# Effect of thermal annealing on the heat transfer properties of reduced graphite oxide flakes: a nanoscale characterization via scanning thermal microscopy

M. Tortello[a], S. Colonna[b], M. Bernal[b], J. Gomez[c], M. Pavese[a], C. Novara[a], F. Giorgis[a], M. Maggio[d], G. Guerra[d], G. Saracco[e], R.S. Gonnelli[a], A. Fina[b,*]

[a]Dipartimento di Scienza Applicata e Tecnologia, Politecnico di Torino, 10129 Torino, Italy

[b]Dipartimento di Scienza Applicata e Tecnologia, Politecnico di Torino, 15121 Alessandria, Italy

[c]AVANZARE Innovacion Tecnologica S.L., 26370 Navarrete (La Rioja), Spain

[d]Università di Salerno, 84084 Fisciano, Italy

[e] Istituto Italiano di Tecnologia, 10129 Torino, Italy

*Corresponding author. Email: alberto.fina@polito.it (Alberto Fina)

**Abstract**

This paper reports on the thermal properties of reduced graphite oxide (RGO) flakes, studied by means of scanning thermal microscopy (SThM). This technique was demonstrated to allow thermal characterization of the flakes with a spatial resolution of the order of a few tens of nanometers, while recording nanoscale topography at the same time. Several individual RGO flakes were analyzed by SThM, both as obtained after conventional thermal reduction and after a subsequent annealing at 1700°C. Significant differences in the thermal maps were observed between pristine and annealed flakes, reflecting higher heat dissipation on annealed RGO flakes compared with pristine ones. This result was correlated with the reduction of RGO structure defectiveness. In particular, a substantial reduction of oxidized groups and $sp^3$ carbons upon annealing was proven by X-ray photoelectron and Raman spectroscopies, while the increase of crystalline order was demonstrated by X-ray diffraction, in terms of higher correlation lengths both along and perpendicular to the graphene planes. Results presented in this paper provide experimental evidence for the qualitative correlation between the defectiveness of graphene-related materials and their thermal conductivity, which is clearly crucial for the exploitation of these materials into thermally conductive nanocomposites.

1. Introduction

The extraordinary electrical, thermal and mechanical properties of graphene [1,2,3] attracted an extremely wide attention on this material. However, the exploitation of graphene in materials for large scale application is still very much limited by the amount and quality of graphene available. In fact, for large scale bulk applications, few-layer graphene, multilayer graphene or graphite nanoplates are typically used and sometimes misreferred to as graphene, while a more rigorous nomenclature should be used [4].

As the properties of graphene-based materials change, sometimes dramatically, when the number of layers increases [5,6] and/or in the presence of defects [7,8,9,10] it is of utmost importance to measure their effective properties instead of assuming those reported in the literature for graphene with potentially very different features.

In particular, in the case of the thermal conductivity of graphene-based materials, the structural features, including nanoplate size, thickness, defectiveness and grain size are known to affect the intrinsic conductivity, as previously reported also for carbon nanotubes [11].

Thermal conductivity for suspended graphene was reported in the wide range of about 2000 to 5000 W $m^{-1}$ $K^{-1}$ [6]. Despite the differences in the phonon modes of carbon nanotubes (CNTs) and graphene, similar mean free path for phonons were reported. The phonon mean free path of CNT was indeed estimated as about 750 nm at room temperature [12], while about 775 nm was reported for suspended graphene [13]. Depending on the flake length/lateral size, this can correspond to a ballistic or close-to-ballistic conduction of the phonons, explaining the exciting conductivity values mentioned above. The conductivity values were reported to be a function of the number of graphene layers, with the highest values observed for a single layer and rapidly decreasing in few-layer graphene. Furthermore, the number of topological defects, re-hybridization defects, incomplete bonding and other defects, as well as the presence of impurities are expected to have an impact on the effective thermal conductivity. To the best of our knowledge, no experimental studies before the present paper were reported on the correlation between the thermal conductivity and the defectiveness of individual flakes of graphene-related materials. However, the conductivity for defective graphene was estimated by molecular dynamics, showing about 50% reduction in the presence of 0.25% defect concentration [7].

The correlation between the chemical defectiveness of graphene and its thermal conductivity was also studied indirectly on nanopapers made of GO [14] or RGO [15] thermally treated at variable temperatures. It has to be noted that measurements on nanopapers reflect not only the properties of flakes constituting the nanopaper but also the quality of contacts between them as well as the density of the nanopaper. However, based on the large increase in the nanopaper thermal conductivity when evolving from GO to RGO [14] or after annealing the RGO nanopaper at high temperature [15] strong beneficial effect of functional groups removal and defectiveness reduction were claimed and a considerable anisotropic thermal conduction of RGO was also clearly observed.

Furthermore, when graphene-based materials are interacting with the surrounding environment, e.g. when supported by a substrate or included as a filler in a polymer matrix, the vibrational properties of the flake may vary significantly. As an example, the thermal conductivity of graphene was reported to decreases from about 2000-5000 W m$^{-1}$ K$^{-1}$ when it is in the exfoliated form and suspended [13,16] to approximately 600 W m$^{-1}$ K$^{-1}$ when it is supported by a substrate like SiO$_2$ [17]. This decrease is most probably explained by the coupling of the out-of-plane acoustic phonons (ZA) to the substrate [17] or to the other graphene layers, together with increased phonon scattering [18].

In graphene-based polymer nanocomposites, the calculated phonon density of states coupling between the polymer matrix and the fillers resulted in a considerable damping of phonons at frequencies lower than 13 THz, with subsequent reduction in thermal conductivity in the range of 30 to 50% compared to uncoupled graphene [19].

Despite the variability described above, graphene-based materials remain of utmost interest as thermally conductive fillers as, unlike other 3D materials like Si-on-insulator whose thermal conductivity strongly decreases with thickness [20], graphene retains a rather high conductivity even when supported or when it is in the multilayer form, as compared to traditional conductive fillers like metals (up to 400 W m$^{-1}$ K$^{-1}$) and conventional graphite (100-400 W m$^{-1}$ K$^{-1}$). Therefore, a lot of effort is focused on the production, possibly on a large scale, of multilayer graphene or graphite nanoplates powders for the exploitation in thermally conductive nanocomposites.

In this regard the possibility to measure, with a nanometric resolution, the thermal properties on the single flakes as a function of the structural defectiveness, size and substrate material may give invaluable information for improving or tuning their properties or just for choosing the most suitable graphene-based material for a particular desired application.

The thermal conductivity of graphene and multilayer graphene can be measured by using Raman optothermal or electrical techniques [6] and graphene and graphite oxide nanoplates have also been studied by means of the thermal flash technique [21].

Scanning Thermal Microscopy (SThM) [22,23,24] also represents a very interesting technique for investigating the thermal properties of materials. In particular, the possibility to employ a new generation of probes that integrate a resistive heater/sensor in a scanning tip with a nanometric curvature radius opens up the possibility for the thermal characterization at the nanoscale [25]. Moreover, with respect to the previously mentioned techniques, SThM can be simpler, faster and can achieve a spatial resolution of the order of a few tens of nanometers. This is much better than what can be achieved with optical techniques, where the resolution is limited by the wavelength due to the diffraction limit. Finally, SThM can also *at the same time* record the topography and the frictional properties (Lateral Force Microscopy) of the investigated material.

In this paper, reduced graphite oxide flakes were produced and thoroughly characterized both in terms of chemical/physical properties and in terms of thermal properties of individual flakes. The same flakes were annealed at 1700°C in vacuum for 1 h to reduce the defectiveness of the $sp^2$ structure and the resulting material was compared to the pristine one. The evolution of the structure was followed by spectroscopic techniques and thermogravimetry, while the analysis of the thermal properties was carried out by Scanning Thermal Microscopy. Both as received and annealed individual flakes, deposited on different substrates, were investigated to experimentally assess the changes in the thermal conductivity after a reduction of their defectiveness.

## 2. Experimental

### *2.1 Graphite nanoplates preparation*

Graphite oxide is prepared using a modified Hummers method [26] starting from 250 g of natural graphite (600 μm average lateral size). The reaction temperature inside the reactor was kept between 0 and 4°C during the addition of oxidants (48h; 15,6 Kg $H_2SO_4$ 98%; $NaNO_3$, 190 g; $KMnO_4$ 1200 g). After that, the temperature of the solution was slowly increased to 20°C and maintained constant for 5 days. The excess of $MnO_4^-$ was removed by reaction with a $H_2O_2$ solution (50 l $H_2O$; 750 g $H_2O_2$ 30%) for 24 hours. After sedimentation, the solution was washed in a mechanically agitated HCl 4%wt solution for 8 h (600:1 washing solution: graphite). The solid was filtered off, washed with osmotic water and dried at 80ºC.

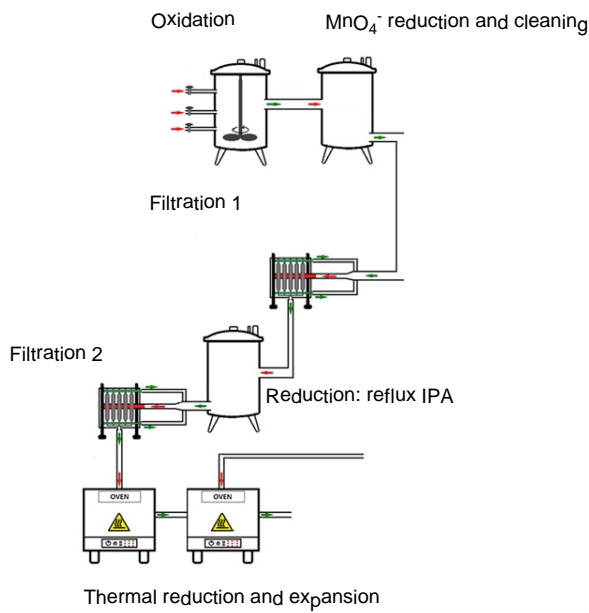

Figure 1: Sketch showing the steps involved in the preparation process of the reduced graphite oxide nanoplates studied in this work and referred to as "RGO".

The chemically-reduced graphite oxide (C-RGO) was prepared in the following way: 100 g of GO were double ultrasonicated in isopropanol and placed at reflux overnight. The solid was then removed, filtered off and air-dried.

Finally, the C-RGO was placed in 2 ovens under Ar atmosphere at 1000 °C for 30 sec for the thermal expansion and then at 1150 °C for 20 min, eventually obtaining a black solid with 0,002 g/l of apparent density. A schematic diagram of the various process steps is reported in Figure 1. From now on, we will name the graphite nanoplates thus obtained as "RGO".

Part of the platelets were also subsequently annealed inside a pure graphite box, to avoid dispersion of graphene powder inside the oven, at 1700 °C for 1 hour at 50 Pa in a graphite resistors oven (manufacturer Pro.Ba., Italy), with heating and cooling rate of 5 °C/min, in order to decrease the amount of defects in the $sp^2$ structure. We will be referring to this material as "RGO_1700 ".

The RGO flakes were deposited, by using the Nitto tape, on two different substrates, namely 285 nm $SiO_2$ on 500 μm Silicon and 100 μm thick PET films.

*2.2 Characterization methods*

X-ray Photoelectron Spectroscopy (XPS) was performed on a VersaProbe5000 Physical Electronics X-ray photoelectron spectrometer with a monochromatic Al source and a hemispherical analyzer. Survey scans as well as high resolution spectra were recorded with a spot size of 100 μm. RGO powder was fixed on a tape and kept in vacuum for about 15 hours prior to the measurement to remove adsorbed volatiles molecules. Data fitting was carried out with XPS Peak 4.1 software following the procedure described in the results and discussion section.

Raman spectra were collected with a Renishaw inVia Reflex (Renishaw PLC, United Kingdom) microRaman spectrophotometer equipped with a cooled charge-coupled device camera. RGO and RGO_1700 flakes, deposited on $SiO_2$/Si using a Nitto tape, were excited with a diode laser source (514.5 nm) with a power density of ~40 W/mm$^2$ at the focus of a 50X microscope objective. The spectral resolution and integration time were 3 cm$^{-1}$ and 100 s, respectively.

Thermogravimetry was performed with a TA Q500 thermobalance, using samples of about 2 mg in alumina open pans heated from 50 °C to 800 °C with a rate of 10 °C/min under air gas flow of 60 mL/min. Onset temperature for volatilization is defined as the temperature corresponding to 3% weight loss.

Morphological characterization of nanoparticles was performed by a high resolution Field Emission Scanning Electron Microscope (FESEM, ZEISS MERLIN 4248). RGO flakes, adhered on a tape, were directly observed without any further preparation.

Wide-angle X-ray diffraction (WAXD) patterns were obtained by an automatic Bruker D8 Advance diffractometer, in reflection, at 35 KV and 40 mA, using the nickel filtered Cu-Kα radiation (1.5418 Å). The instrumental broadening (*β*inst) was determined by fitting of Lorentzian function to line profiles of a standard silicon powder 325 mesh (99%). For each observed reflection, the corrected integral breadths were determined by subtracting the instrumental broadening of the closest silicon reflection from the observed integral breadths, $\beta = \beta_{obs} - \beta_{inst}$. The correlation lengths ($D$) were determined using Scherrer's equation.

$$D = \frac{K\lambda}{\beta \cos\theta} \tag{1}$$

where $\lambda$ is the wavelength of the incident X-rays and $\theta$ the diffraction angle, assuming the Scherrer constant $K = 1$. For the evaluation of amourphous fraction, WAXD spectra for calcined petroleum coke supplied by Asbury Graphite Mills Inc was used as a reference.

## 2.3 Atomic Force and Scanning Thermal Microscopy

The topography, lateral force and scanning thermal maps were acquired by using an Innova® AFM from Bruker Corp. equipped with an SThM tool (VITA module). The VITA-HE-GLA-1 contact-mode probes were adopted which feature a resonant frequency of about 50 KHz and a spring constant of 0.5 N/m. Furthermore, they incorporate a thin Pd film deposited near the apex of the $Si_3N_4$ probe that functions at the same time as a heater and temperature sensor. The heater/sensor, which works as the resistive element of a Wheatstone bridge, is then Joule-heated and the bridge voltage is monitored as the tip is scanned over the sample. The voltage is then converted to temperature (as explained below) and if the heat flow between the tip and the sample increases (more thermally conducting regions) the sensor temperature decreases while the opposite situation occurs for less conducting areas. In order to convert the voltage to heater temperature the following procedure was adopted: i) Before scanning, the resistance of the sensor is first measured by a digital multimeter at very small currents, in order to avoid Joule heating. Typically, the resistance of the probe is of the order of 300 Ω. ii) While scanning, the probe is Joule-heated and the bridge voltage is measured. The actual resistance of the heater is calculated by using the standard Wheatstone bridge formulas including, besides specific parameters of the electronics, the resistance previously measured in the absence of heating. iii) The resistance is converted to temperature following a procedure described by Tovee *et al*. [27].

## 3. Results and discussion

### 3.1 RGO nanoplates characterization and evolution upon annealing

Scanning electron microscopy was used to characterize the RGO flakes before and after annealing. In untreated RGO, a highly expanded accordion-like structure is visible (Figure 2a,b), consisting of randomly aggregated folded sheets in a three dimensional porous architecture. Despite accurate measurements of the thickness of the fakes are not possible in this conditions, flakes thickness is estimated to be in the range of a few nm. The annealing process at 1700°C does not significantly alter the morphology, structure or apparent thickness of the graphite plates, even if the structure appears slightly more expanded after the annealing treatment (Figure 2c,d).

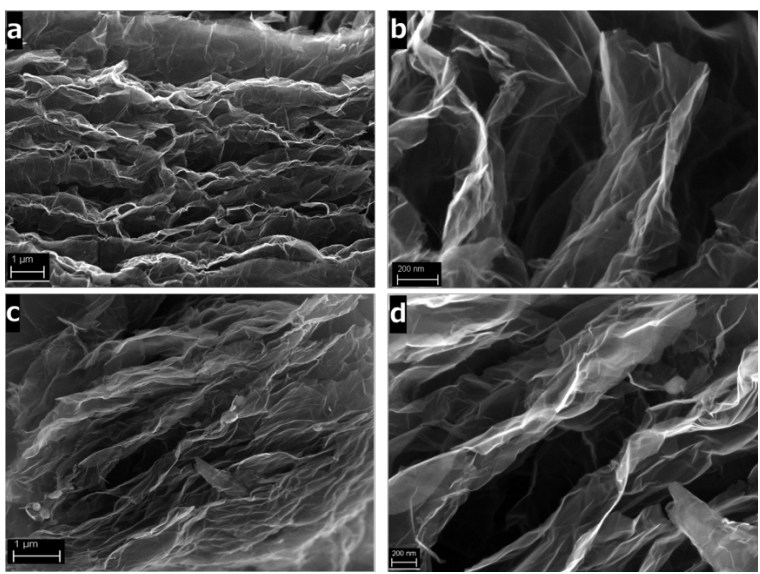

Figure 2: (a) FESEM image of the RGO nanoplates. (b) The same as in (a) but with a higher magnification; (c) FESEM image of the thermally treated RGO nanoplates (RGO_1700). (d) the same as in (c) but with a higher magnification. The length scale is 1 μm for (a) and (c) and 200 nm for (b) and (d).

Composition of the flakes before and after annealing was studied by XPS analyses which evidenced the presence of 3.2 atomic % oxygen in pristine RGO, while most of the oxidized groups were eliminated during the annealing, as the oxygen concentration has reduced to 0.4 atomic % after annealing. However, it is worth noting that the treatment does not eliminate only the oxygen as the weight loss upon treatment was measured to be in the range of 10% wt. This is partially explained by the fact that oxygen is typically eliminated as CO or $CO_2$ [28], thus removing some carbon which contributes to the weight loss. Furthermore, elimination of other unstable carbon in proximity of the edges and/or next to surface defects is also possible, contributing to the total weight loss during the annealing. To further investigate the materials chemical structure before and after the annealing, XPS spectra were carefully fit on both $O_{1s}$ and $C_{1s}$ signals, as shown in Figure 3.

For $O_{1s}$ in RGO, despite the relatively low intensity of the signal due to low oxygen content, two partially overlapping components were clearly observed. Deconvolution of this peak with two Gaussian-Lorentzian (80:20) after Shirley background subtraction delivered a satisfactory fitting with peaks at 533.2 and 531.0 eV, the higher binding energy showing a slightly higher intensity. The signal at 533.2 eV is generally assigned to single-bonded oxygen, C-O, whereas the peak at lower binding energy (531.0 eV) is typically assigned to double-bonded oxygen species, namely carbonyl group C=O or carboxylic groups O=C-OH [29,30]. Despite multiple peaks would be expected at slightly different positions for both single-bonded oxygen (e.g. C-O-H and C-O-C groups), as well as for double-bonded oxygen (e.g. C=O and O=C-OH groups), taking into account the relatively low signal for O1s, reliable fitting with multiple peaks signals seemed unrealistic and a simpler deconvolution was preferred in this paper. In fact, the existence of both single- and double-bonded oxygen species revealed the presence of different types of residual oxidized species on RGO, which is consistent with previously proposed thermal evolution of graphene oxide [31].

On the other hand, annealed RGO showed a very weak $O_{1s}$ band which was fitted by a single Gaussian-Lorentzian with peak at 532.4 eV. The lower intensity clearly reflects the total amount of oxygen, which was strongly reduced during the annealing at 1700°C, whereas the position of the peak is quite close to the one assigned to single-bonded oxygen species. This suggests that carbonyl and carboxylic groups are mostly eliminated during annealing, whereas some C-O groups are still observed after the high temperature annealing, likely in the form of phenolic groups, as previously proposed by Ganguly at el. [31].

Carbon ($C_{1s}$) band for RGO showed a maximum at 284.5 eV and a strongly asymmetric shape extending to higher binding energies up to about 295 eV, clearly showing the overlapping of several different signals related to carbon atoms with different chemical environments. This binding energy distribution is typical for reduced graphene oxide and fits with several peaks have been reported by many different authors [29,30,31,32,33]. However, owing to the extensive signals overlap, the interpretation of XPS deconvolution with several peaks can be rather subjective and assignment of the different peaks is often controversial in literature. While the main peak at 284.5 eV is certainly assigned to C-C in $sp^2$ configuration, other peaks typically used to fit RGO include C=O (287.5-287.7 eV [a]), COOH (288.5-290.1 eV[a]), C-OH (285.3-286.0 eV[a]), C-O-C (285.5-286.5 eV[a]), C-C in $sp^3$ configuration (285.5-286.3 eV[a]), generic defects (285.3-285.5 eV[a]) and π-π* shake-up transition, typically indicated as a very broad band centered at about 291 eV. In the present paper, fitting was carried out by fixing the positions of $sp^2$ C-C (284.5 eV) and π-π* shake-up transition at 291 eV, while leaving two additional peak positions free for fitting optimization. The addition of more than two peaks for oxidized species did not deliver better fits of the experimental curve. With the above conditions, the optimized positions obtained for the two peaks by the fitting procedure were at 285.5 and 287.3 eV, with relative intensity of approx. 1.8:1. Comparing these results with literature assignment, we can reasonably conclude that the lower binding energy peak has to be assigned to C-OH and/or C-O-C groups, whereas the higher binding energy peak is representative of C=O and/or COOH chemical functions.

The same fitting procedure was applied to RGO_1700 and the spectral position of the two peaks turned out to be almost identical to the case of RGO described above, with relative intensity of 2.1:1. Comparing the deconvoluted signals for RGO and RGO_1700, we can notice that, besides the expected decrease of intensities for the peaks of the oxidized species after annealing, some differences are present for the peaks related to graphitic C-C. In particular, a decrease from 0.84 to 0.71 for the full width at half maximum (FWHM) of $sp^2$ C was found upon annealing. The FWHM was previously reported to be sensitive to heterogeneity of both the chemical and the structural environment of carbon [29]. Furthermore, the intensity of the π-π* shake-up band was almost doubled in intensity upon annealing, which was also previously assigned to a restoration of aromaticity in the structure after removal of oxygen [29].

---

[a] Peak position interval is given taking into account fitting proposed by the authors cited above

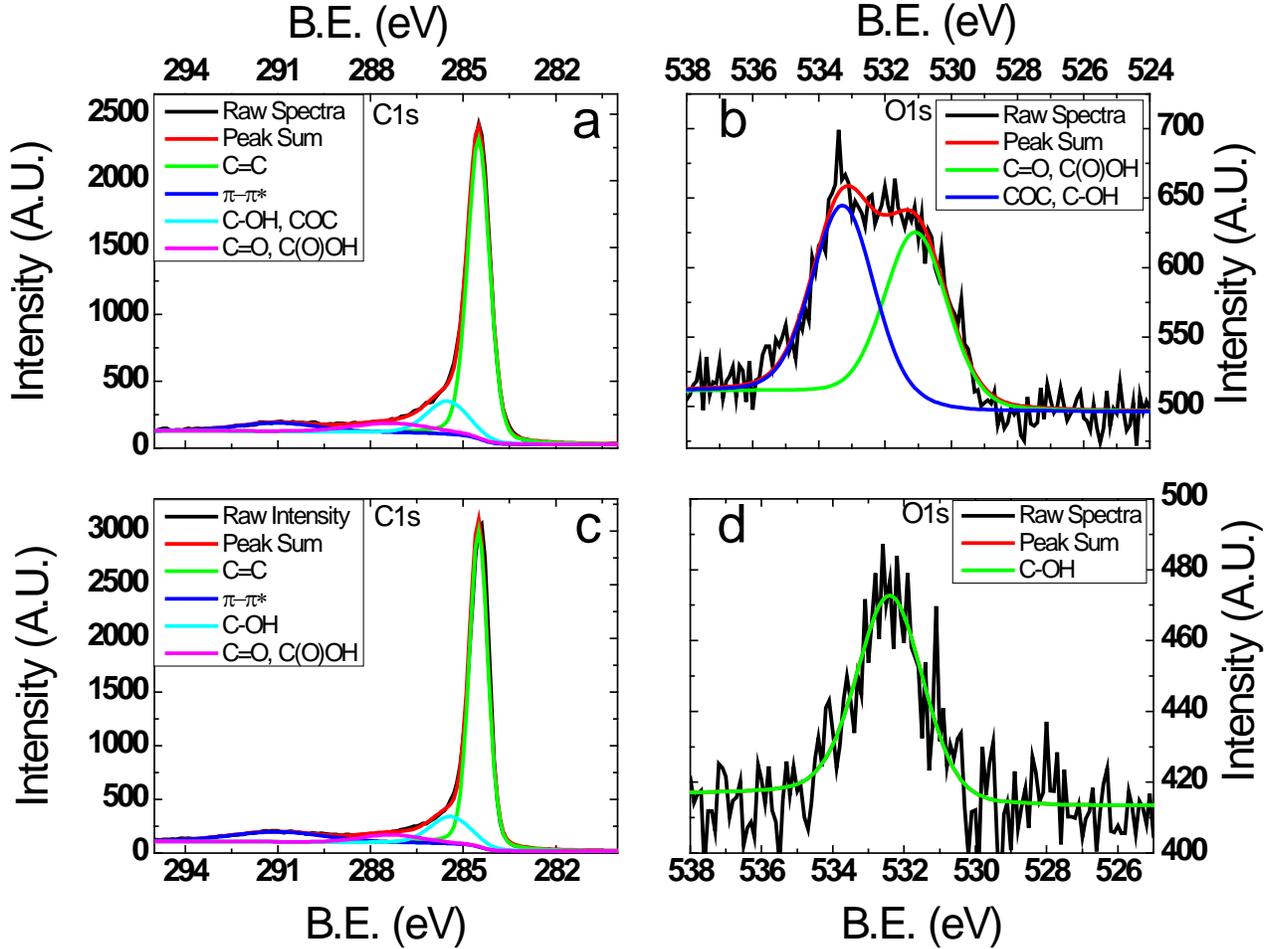

Figure 3. XPS curves with their deconvolution peaks for C1s and O1s of RGO (a, b) and RGO_1700 (c, d), respectively.

Raman spectroscopy has been performed on RGO and RGO_1700 in order to extract information about the samples microstructure. Figure 4 shows first- and second-order Raman spectra of representative nanoplates which present a noticeable variation after thermal annealing (Raman analyses conducted on several nanoplates replica show almost identical vibrational features). The first-order RGO Raman spectrum is composed by two strong resonances at ~1350 cm$^{-1}$ (defect-related D band) and at ~1583.5 cm$^{-1}$ (G band); the last vibrational mode is convolved with a weaker feature at ~1615 cm$^{-1}$ which can be ascribed to the disorder-induced D' band [34]. The second-order RGO spectrum shows three different bands (G´, D + G, and 2D´) typical of disordered graphitic materials and oxidized graphene [35]. The G' mode at ~2692 cm$^{-1}$ is due to a double resonance intervalley Raman scattering process with two iTO phonons at the K point, whose intensity is sensitive to the presence of structural disorder.

The D+G peak at ~2927 cm$^{-1}$ is a combination mode and the 2D' peak at ~3194 cm$^{-1}$ is the second-order mode of the D' band. After annealing, the Raman spectrum yields, as expected, a marked evolution towards a more ordered graphitic structure. Actually, the G band evidenced at ~1575 cm$^{-1}$ is shrunk and much more intense, the D' mode disappears and the D peak, still present at ~1352 cm$^{-1}$, becomes very weak. In particular, the $I_D/I_G$ ratio strongly decreases from ≈ 1.19 for the non-annealed sample (suggesting a high concentration of defects in the aromatic structure, including sp$^3$ carbons and possible residual oxidized groups [36,37,38]), to ≈ 0.023 for the thermally-treated one. Coherently with the expected decrease of the structural disorder, the second-order spectrum of RGO_1700 shows a higher intensity for the G' peak located at about 2705 cm$^{-1}$. Such band shows a certain asymmetry, which can be justified with the interplanar stacking order of graphene layers. On one hand, graphitic materials with high staking order such as highly oriented pyrolytic graphite (HOPG) exhibits an asymmetric lineshape that can be fitted with two Lorentzian components (G´$_{3DA}$ and G´$_{3DB}$) due to Bernal stacking arrangement of individual graphene layers [34]. On the other hand, materials characterized by graphene layers randomly oriented (turbostratic stacking) are featured by a symmetric G' band that can be fitted with a single Lorentzian component (G'$_{2D}$) [29]. As it can be noticed in the inset of Figure 4, the G' band of RGO_1700 specimen shows the coexistence of G'$_{2D}$, G´$_{3DA}$ and G´$_{3DB}$ Raman peaks, with a fractional 3D graphitic volume experimentally found as $I_{G'3DB}/(I_{G'3DB} + I_{G'2D})$ ≈ 60%, giving a proof of a high degree of three-dimensional order.

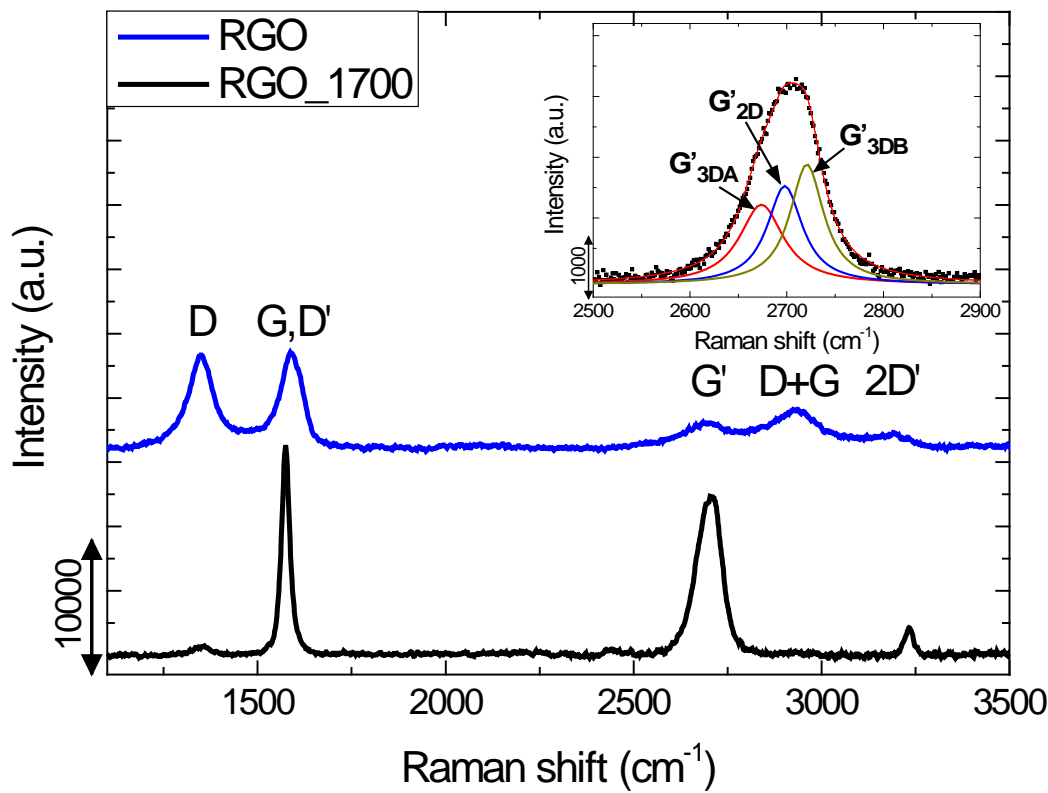

Figure 4: Raman spectra for the RGO nanoplates (blue line) and RGO_1700 ones (black), thermally annealed in vacuum at 1700 °C for 1 h. The background due to photoluminescence emission has been subtracted. The inset shows the G' Raman band spectrum for RGO_1700 nanoplates fitted with 3 Lorentzian components (G'$_{2D}$, G´$_{3DA}$ and G´$_{3DB}$) as discussed in the text.

X-ray diffraction was used to study the crystalline order of RGO and annealed RGO; WAXD patterns for both materials are reported in Figure 5.

For RGO, the 002 reflection at $2\theta = 26.3°$, corresponding to the spacing between graphitic layers (0.339 nm), exhibits a half height width $\beta = 1.35°$, giving an average correlation length perpendicular to the graphitic layers of $D_{002} = 6.7$ nm (Figure 5, curve B). This 002 reflection is overlapped to a significant amorphous halo, which is similar to the main diffraction halo of disordered graphitic materials, e.g. calcined petroleum coke (Figure 5, curve A). A deconvolution of the 002 peak and of the amorphous halo in the $2\theta$ range 18°-40° allows evaluating the amount of graphite nanoplatelets close to 30 wt% of the RGO sample. This confirms a significant disorder in the RGO structure, in agreement with Raman results. The WAXD pattern of RGO also shows a less intense amorphous halo, centered at $2\theta = 43.5°$, essentially similar to the second amorphous halo of petroleum coke (Figure 5, curve A). The WAXD pattern of RGO also shows weak reflections at $2\theta = 54.4°$ and $77.5°$, corresponding to 004 and 110 graphite reflections. It is worth adding that the half height width of the isolated 110 reflection ($\beta = 0.63°$), allows evaluating an average correlation length parallel to the graphitic layers (more precisely perpendicular to the 110 planes) of $D_{110} = 18$ nm. Moreover, the absence of 101 and 112 reflections clearly indicate a complete absence of hexagonal order and hence the formation of a turbostratic structure.

After the high temperature thermal annealing, significant modification of the XRD pattern, with all reflections becoming sharper and more intense, was observed. A narrowing of the *002* reflection was obtained after annealing, with half height width $\beta = 0.98°$, corresponding to an increase of the average correlation length perpendicular to the graphitic layers up to $D_{002} = 9.3$ nm. The peak deconvolution in the $2\theta$ range 18°-40° indicates that the amount of graphite nanoplatelets increases up to 52 wt% of the sample. At higher $2\theta$ angles, the amorphous halo between 42 and 46° definitely changes with the appearance of the in-plane 100 reflection ($2\theta = 42.5°$; $d = 0.213$ nm) but also with its centering at $d=0.205$, corresponding to the position of the 101 reflection. This indicates that annealing also brings to some partial hexagonal order in the stacking of the graphitic layers. The half height width of the isolated 110 reflection ($\beta = 0.47°$), indicates an increase of the average correlation length parallel to the graphitic layers up to $D_{110} = 24$ nm.

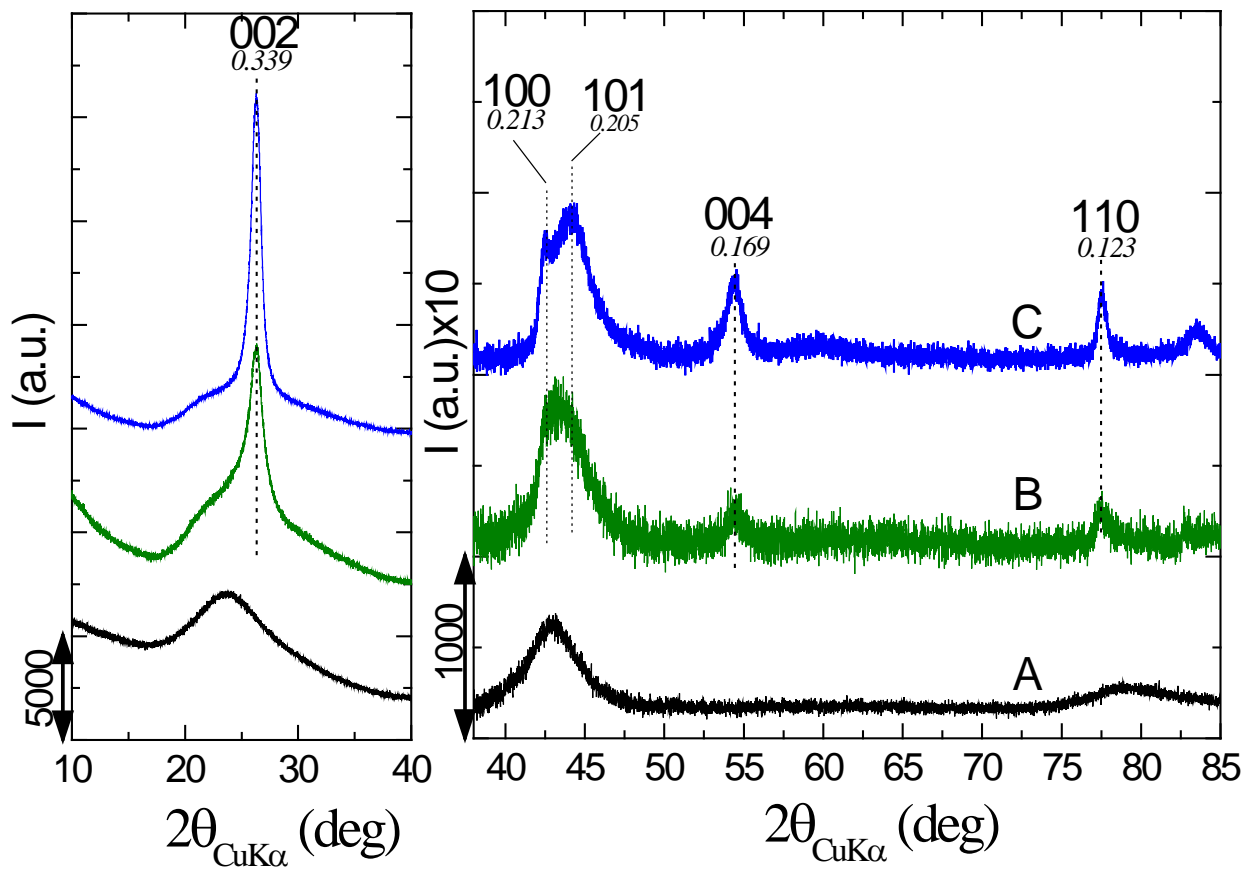

Figure 5. X-ray diffraction patterns (CuKα) of RGO (B) and annealed RGO (C). The amorphous fraction of the samples presents diffraction halos similar to those of calcined petroleum coke (A).

Thermogravimetry was also used to indirectly assess the defectiveness of RGO. Indeed, large and well graphitized flakes are typically stable in air up to temperatures in the range of 700°C while the presence of $sp^3$ carbons as defects on the surface or at the flake edges may trigger the subsequent oxidation and volatilization of graphene-based materials [39]. Pristine RGO showed an onset of weight loss at 558°C, after which the volatilization process continues, leading to a stable residue of about 7% above 750°C (Figure 6), mainly corresponding to silicate impurities in the natural graphite used, as indicated by the energy dispersive spectroscopy analysis under SEM microprobe. It is worth noting that volatilization occurs in two partially overlapping steps, as evidenced by the shape of the mass loss rate plot, showing a main peak at 643°C and a shoulder centered at about 695°C, which can be explained by the polydispersity of lateral flake size. On the other hand, annealed RGO showed a significantly higher thermoxidative stability, as evidenced by the onset temperature at 748°C, further supporting the reduction of $sp^3$ carbons upon annealing.

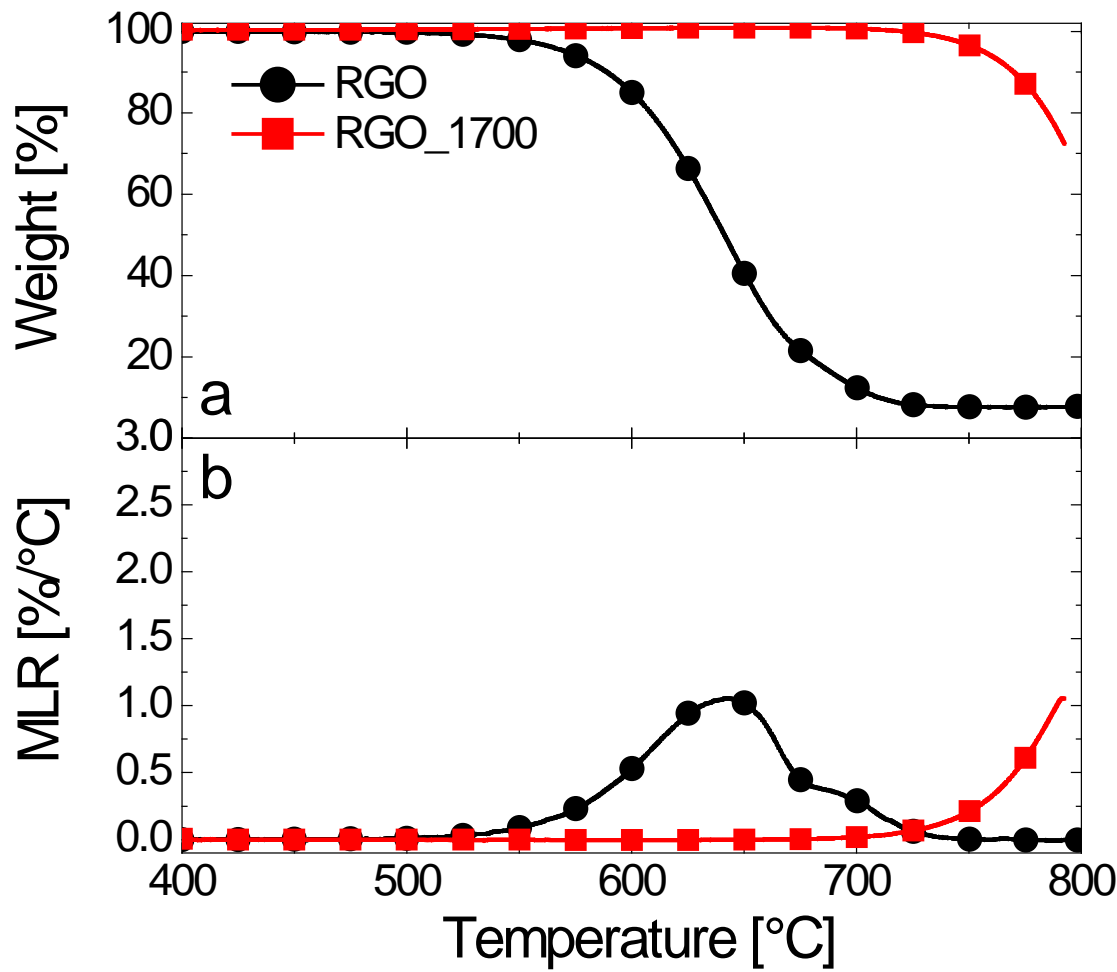

Figure 6: (a) Thermogravimetric curve of the RGO (black line) and RGO_1700 (red line) samples. (b) The first derivative of the curves shown in panel (a).

Based on the characterization results described in this section, we can conclude that a dramatic evolution of the RGO chemical structure and morphology was obtained during the thermal annealing process at 1700°C. In particular, a strong reduction of oxygen content was evidenced by XPS analysis, especially by elimination of carboxylic and carbonyl groups, as the few remaining oxygenated groups can be assigned to phenols. Elimination of oxygen from RGO was reported to introduce vacancies in the graphene flakes, for which extensive annealing was reported to require temperatures higher than 1700°C [29]. This suggests that a full aromatization cannot be expected by the treatment applied in this paper. However, a significant reduction of defectiveness was evidenced by Raman, together with an evolution towards a more ordered stacking of graphene layers. This was also confirmed by XRD, which evidenced ordering of the structure both along and perpendicular to the graphene planes.
The effect of such reduction of the structural defectiveness on the thermal properties of individual flakes is described in the following section.

*3.2 AFM/SThM results and discussion*

Scanning thermal microscopy with nanometric resolution was used to investigate the thermal properties of the RGO flakes. Indeed, the heated SThM probe imposes a heat transfer from the tip through the material and eventually to the substrate, which allows to locally assess the thermal properties of the sample scanned by the probe.

Let us first recall that the thermal resistance is defined as $R = \frac{\Delta T}{\dot{Q}}$, where $\Delta T$ is the temperature difference and $\dot{Q}$ is the heat flux between the hot and cold regions. Then, when applying this general definition to the SThM system, the hot and cold regions correspond to the probe heater temperature and ambient temperature, respectively. If we just limit ourselves to a qualitative comparison between pristine and annealed flakes supported by the same substrate, the total thermal resistance seen from the probe can be modeled by using a lumped parameters electronic analogue [27, 40]:

$\frac{1}{R_{tot}} = \frac{1}{R_{air}} + \frac{1}{R_c} + \frac{1}{R_{h-t} + R_{t-s} + R_s}$, where $R_{air}$ is the thermal resistance from the heater to the surrounding ambient air, $R_c$ is that to the cantilever base, $R_{h-t}$ is the thermal resistance between the heater and the tip apex, $R_{t-s}$ is the tip-sample contact resistance and $R_s$ is a thermal resistance related to the thermal properties of the sample. It is worth noting here that in our case $R_s$ includes also the thermal boundary resistance between the flake and the substrate and the spreading resistance of the substrate. The most significant complication to the problem is given by the fact that the thermal conductivity of the RGO flakes is expected to be highly anisotropic with an in-plane conductivity, $k_{ab}$ that can likely be up to two orders of magnitude higher than the out of plane one, $k_c$ as it occurs for instance in HOPG. Furthermore, this high in-plane conductivity could possibly lead to a heat transfer term from the flake surface to the air. Even though some reasonable approximations could be made in order to simplify the problem [41], it is clear that an exact determination of the thermal conductivity of the flake would necessarily require the determination of $R_{t-s}$ (also related to the contact area between the tip apex and the sample) and of the boundary resistance between the RGO flake and the substrate, to be included into a detailed finite element model, which is beyond the scope of this paper. Nevertheless, from the above resistance model we can qualitatively infer that if the thermal conduction of the RGO flake improves (or, equivalently, $R_s$ decreases), $R_{tot}$ decreases accordingly, assuming all the other terms remain substantially unchanged. From the experimental point of view this means that, at constant heating power, a more thermally conducting RGO flake would feature a lower tip heater temperature than a less conducting one when supported by the same substrate. If we use the temperature of the probe when in contact with the substrate, $T_{sub}$, as a fixed reference (as experimentally observed), the above considerations imply that, since in our cases we have always found out that $T_{sub} \geq T_{flake}$, the temperature difference between the substrate and the RGO flake should be higher for more conducting flakes than for less conducting ones.

Moreover, when decreasing the thermal conductivity of the substrate (e.g. for PET it is one order of magnitude lower than for $SiO_2$), we will see that the temperature difference between the flake and the substrate will increase with respect to the case of $Si/SiO_2$. In fact, it can be easily demonstrated that the absolute decrease of the spreading resistance [42] of the substrate that occurs when you enlarge the heat flow area (as it happens when the flake/substrate configuration substitutes the tip/substrate one), is larger when the thermal conductivity of the substrate is smaller. As a consequence, the temperature difference substrate-flake is also expected to be higher.

It is also worth pointing out here that these considerations are reliable and are correctly related to the thermal properties of the materials under study only when the tip is scanned over a rather flat surface (compared to the probe apex height). Indeed, if the topology is highly irregular, the measured temperature can be strongly affected by topological effects [43]. For example, if the tip is being scanned on the top of a rather high "hill", the temperature measured by the sensor will be much higher than the actual temperature of the surface which can thus be overestimated due to the fact that a smaller sample area is heated by the probe, either because it is surrounded by more air (yielding a bad conduction) and/or because the tip-sample contact area decreases with respect to an ideal probed flat region. On the other hand, when the tip is inside a narrow "valley", the actual sample temperature could be underestimated because now the tip is surrounded by the sample (generally more conducting than the air) and/or the tip-sample contact area increases.

When measuring in ambient atmosphere, as it is our case, both heat losses through the cantilever and air are present at the same time. The resulting parallel resistance, $R_{a/c}$, that describes these two effects as a whole can be determined when the tip is off the sample. We determined $R_{a/c} = \frac{R_{air} R_c}{R_{air} + R_c} \cong 1.9 \times 10^5 \ K/W$, when the tip is far off the sample and where the thermal resistances have been determined by taking into account the power generated by the heating/sensing element (Pd resistor) close to the tip. When the tip is approaching the sample, part of the heat flows into the sample through the air, thus decreasing $R_{a/c}$. If $R'_{a/c}$ is the new thermal resistance when the tip is in contact with the sample, it can be determined by performing approach and retract measurements [44] and by recording the temperature of the probe just before contacting the sample (≈ 4-5 nm above) since, as a matter of fact, the resistance immediately before the tip-sample contact can be reasonably identified with $R'_{a/c}$. The value measured in this way is $R'_{a/c} \cong 1.7 \times 10^5 \ K/W$ and, as expected, it slightly changes around this value depending on the substrate material.

When the tip is contacting the substrate, $R'_{a/c}$ is in parallel with the resistance $R''$ describing the heat flow through the tip into the sample. $R''$ can be considered as the sum of the resistance between the heater and the tip apex, the tip-substrate contact resistance and the substrate spreading resistance. We found that $R'' \cong 2.3 \div 3.8 \times 10^7 \ K/W$, depending on the substrate ($SiO_2$ and PET).

We performed AFM and SThM measurements on the RGO and RGO_1700 nanoplates in order to check their morphological and thermal properties, before and after the annealing procedure.

AFM measurements were conducted by acquiring, at the same time, the topographic, lateral force and scanning thermal maps. Several different flakes were observed and measured, showing some variability in both lateral size, ranging between 0.5 and 2 µm, as well as in thickness, ranging from about 4 to 15 nm.

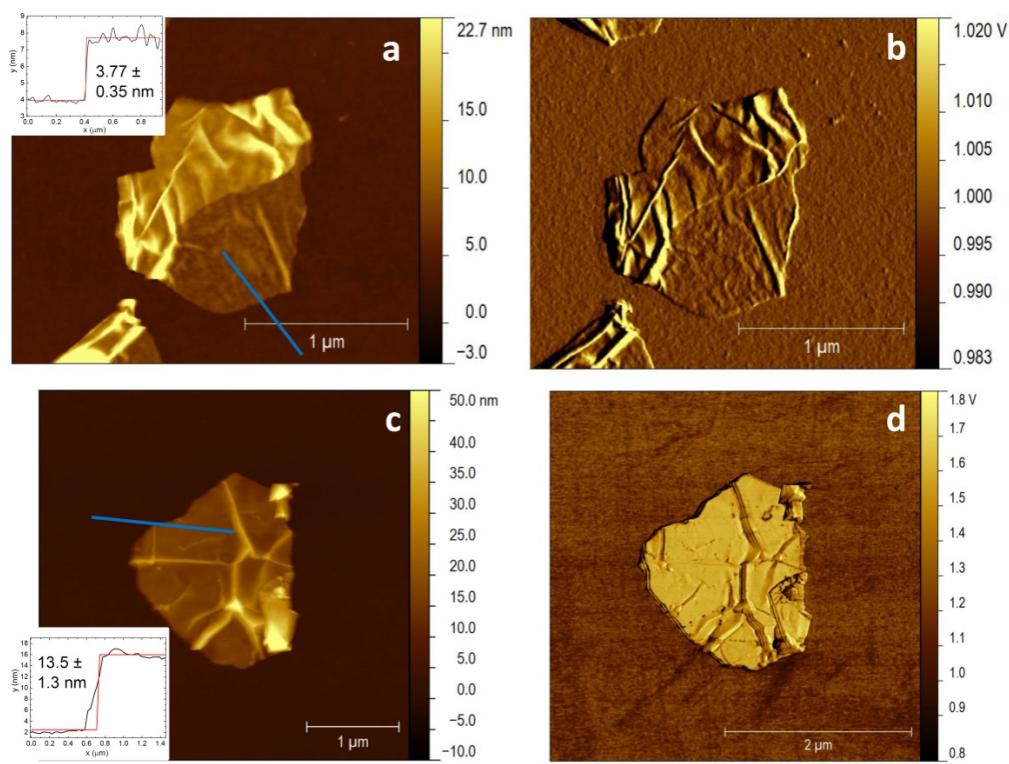

Figure 7: Topography (a) and lateral force (b) maps of RGO nanoplates deposited on a Si/SiO$_2$ substrate. The inset shows the height profile indicated by a blue line in the main panel. (c) and (d): the same as in (a) and (b) but for the RGO_1700 nanoplates.

Figure 7a reports a topographic image of a RGO nanoplate supported by a Si/SiO$_2$ substrate. The lateral size is about 1 μm and its surface features several irregularities such as wrinkles and bulges, which were always observed on this grade of graphite nanoplates and that are mainly caused by the thermal expansion process [45]. However, in the lower right part of the plate it is possible to observe a rather large flat region whose average thickness, as shown in the inset to Figure 7a, is less than 4 nm, one of the lowest observed in this work. Panel b shows the lateral force, or friction imaging, map which is highly (but not exclusively) sensitive to the friction between the tip and the graphite plate (or the tip and the substrate) during the contact mode scanning. Due to the sensitivity of the lateral-force signal to the rapid changes in the slope of the sample topography, this kind of images is also useful, as in this case, to highlight the edges and wrinkles and thus often give a clearer idea of the topography.

Panel c shows the morphology map of a RGO_1700 sample on Si/SiO$_2$ substrate. This plate is larger and thicker (see inset) than the previous one and, although it presents as well several wrinkles and is clearly folded along the lower right edge, it features some flat areas where SThM measurements can be carried out avoiding artifacts due to the sample topology. Panel d reports the lateral force map which very clearly shows all wrinkles, folding and other tiny features of the nanoplate. In agreement with FESEM results, no significant changes in the morphology of the graphite nanoplates have been in general noticed as a consequence of the annealing procedure.

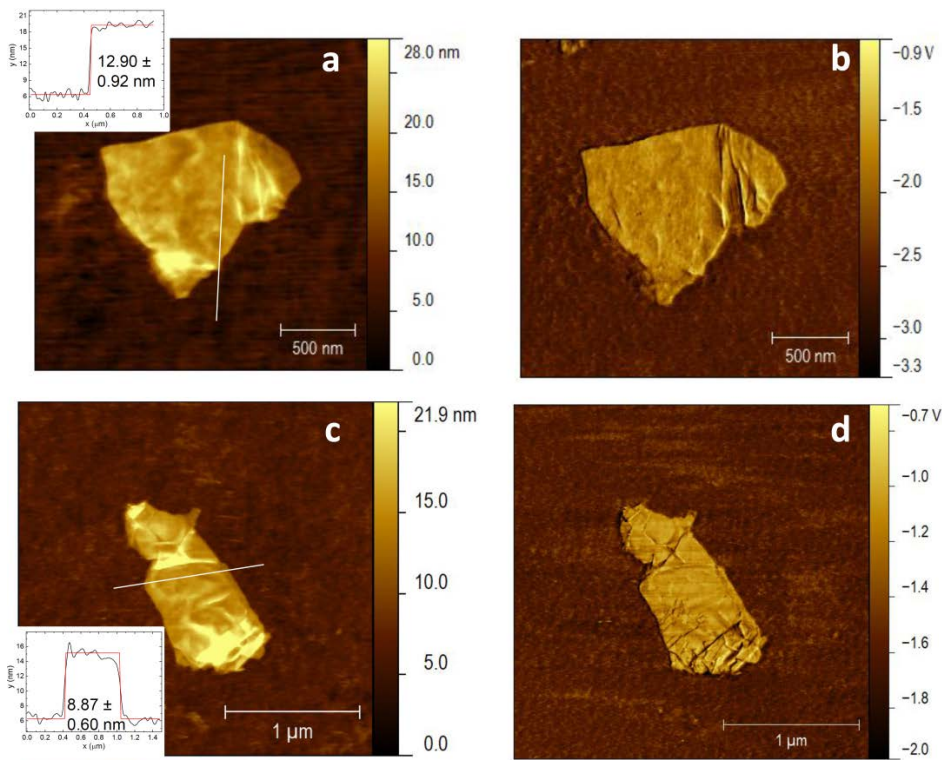

Figure 8: Topography (a) and lateral force (b) maps of RGO nanoplates deposited on a PET substrate. The inset shows the height profile indicated by a white line in the main panel. (c) and (d): the same as in (a) and (b) but for the FLG24_1700 nanoplates.

Figure 8a shows the topographic image of a RGO plate supported by PET. Similarly to the nanoplates on $SiO_2$ reported above, the lateral size here is about 1 μm while the thickness is approximately 13 nm (see inset). Some bulges and wrinkles are present on the right part of the nanoplate, as highlighted by the friction map shown in panel b. Panel c shows the topography of an almost rectangular (about 1 x 0.5 μm$^2$) nanoplate of RGO_1700, about 9 nm thick, that is rather flat at the center but more irregular at the ends, particularly the lower one, as evidenced more clearly by the lateral force map shown in panel d.

These images evidenced that, on both $SiO_2$ and PET substrates, the deposition procedure was successful and led to the adhesion of individual flakes, obtained by mechanical separation of these plates from the expanded structure observed by scanning electron microscopy and shown in Figure 2.

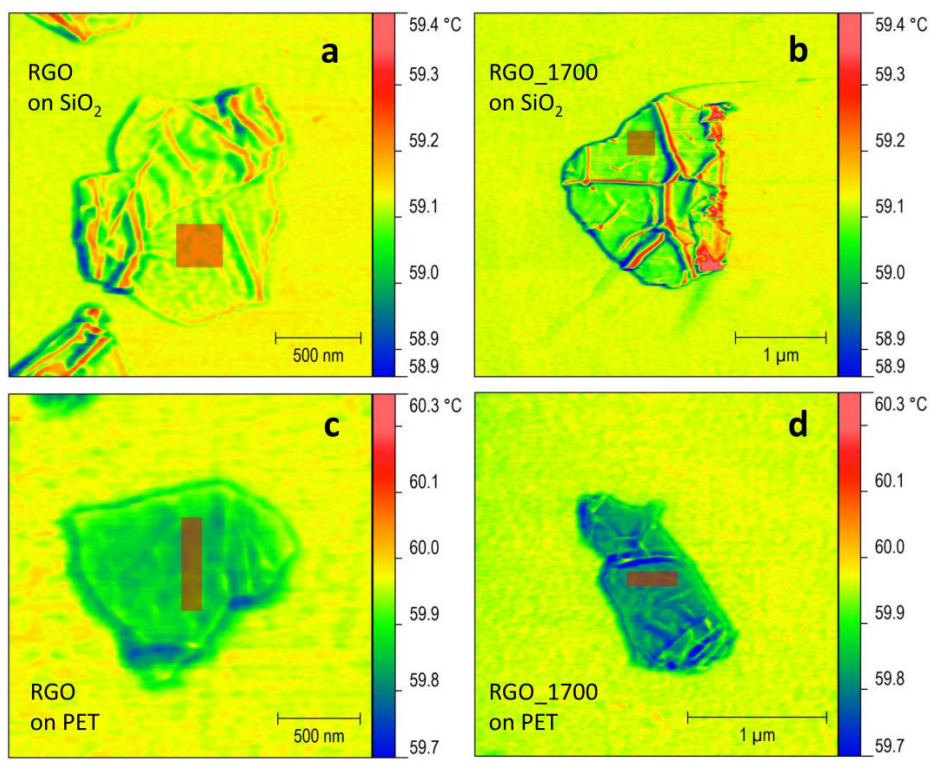

Figure 9: SThM maps of RGO (a) and RGO_1700 (b) deposited on Si/SiO$_2$ and RGO (c) and RGO_1700 (d) deposited on PET. (note that the length scales for flakes on SiO$_2$ and on PET are different)

Figure 9a shows the thermal map of the RGO nanoplate supported by Si/SiO$_2$ measured together with those previously reported in Figure 7a and b. First of all it can be seen that wrinkles, bulges and edges cause a relatively higher temperature variation with respect to the flat areas. This is due to the topological effects discussed above, that occur when the tip is scanned over a "hill" or inside a "valley". On the other hand, focusing on the probe temperature when scanning on the flat areas of the nanoplate, the temperature in those regions looks rather constant at about 59.10 ± 0.05 °C. Furthermore, it is clearly observable from this thermal map that the temperature value on the substrate is very close to the temperature measurement on the RGO flake. This means that the heat dissipation from the SThM probe to the substrate is very similar to the dissipation on the surface of the RGO flake. This does not directly mean that the RGO features the same conductivity as the SiO$_2$/Si substrate but simply that their apparent thermal conductance is close. Indeed, the apparent conductance seen by the tip in contact with the nanoplate takes into account the highly anisotropic thermal conduction of the graphitic systems along with the effect of the underlying substrate, as both terms contribute to the total thermal resistance, as discussed above. By calculating the average temperature on the masked region (whose area is about 0.068 μm$^2$) in Figure 9a and subtracting it from the temperature measured on the substrate (obtained by averaging on a similar area on the substrate), a temperature difference as low as ΔT=7.6±14.2 x 10$^{-3}$ K was obtained. The average temperature is usually calculated over thousands of points (≈ 3300 in the case of panel a), while the uncertainty is obtained by calculating the standard deviation. The uncertainty of the temperature difference is obtained by summing up the uncertainties of the two terms of the subtraction. Given the low value and the relatively large experimental uncertainty, no differences in apparent conductance may be claimed between SiO$_2$/Si substrate and SiO$_2$/Si supported RGO. The same scanning thermal measurements were carried out on RGO_1700 nanoplates on SiO$_2$/Si and results obtained for a representative flake are reported in Figure 9b. Apart from the analogous topological effects on the sharp edges, it is evident that the temperature when scanning on the flat areas of the sample is significantly *lower* than that on the substrate, indicating a higher thermal conductivity of the annealed nanoplate with respect to the RGO flake. The temperature difference of the sensor on the substrate and on the nanoplate (determined as an average of temperatures in the red masked area in panel b) was ΔT=42.2 ± 14.6 x 10$^{-3}$ K which, within the uncertainty, is significantly higher than the value obtained for the pristine RGO sample. Several other nanoplates have been thermally characterized to confirm the results described above. Indeed, for all the flakes, the sensor temperature difference substrate-nanoplate has been found close to zero for the RGO samples and considerably higher for the RGO_1700 ones.

As anticipated above, if similar measurements were conducted on a substrate with a lower thermal conductance, then the observed temperature differences should increase with respect to the Si/SiO$_2$ case. This is what we indeed investigated by using PET as a substrate, whose thermal conductivity is about one order of magnitude smaller than that of bulk SiO$_2$, while the thermal boundary resistance can be assumed to be of the same order of magnitude [6].

Figure 9c reports the thermal imaging map of an RGO graphite nanoplate supported by a PET substrate. Unlike what observed in panel a, a significant temperature difference between the substrate and the sample can be seen, in that case ΔT=103.1±17.7 x 10$^{-3}$ K, clearly much higher than for the Si/SiO$_2$ substrate. This confirms, as expected, how a less thermally conducting substrate does enhance the temperature contrast in the SThM maps.

Panel d shows the SThM map of the RGO_1700 nanoplate supported by PET. Also in this case, the temperature contrast between the sample and the substrate (ΔT=143±13 x 10$^{-3}$ K) is higher than for the pristine RGO on the same substrate. Again, this fact indicates that the thermal conductivity of the RGO flakes is significantly increased after the annealing treatment, which is primarily correlated with the reduction of defects in the sp$^2$ structure, as discussed above. The results reported for individual flakes on PET have been reproducibly observed on several other flakes as well.

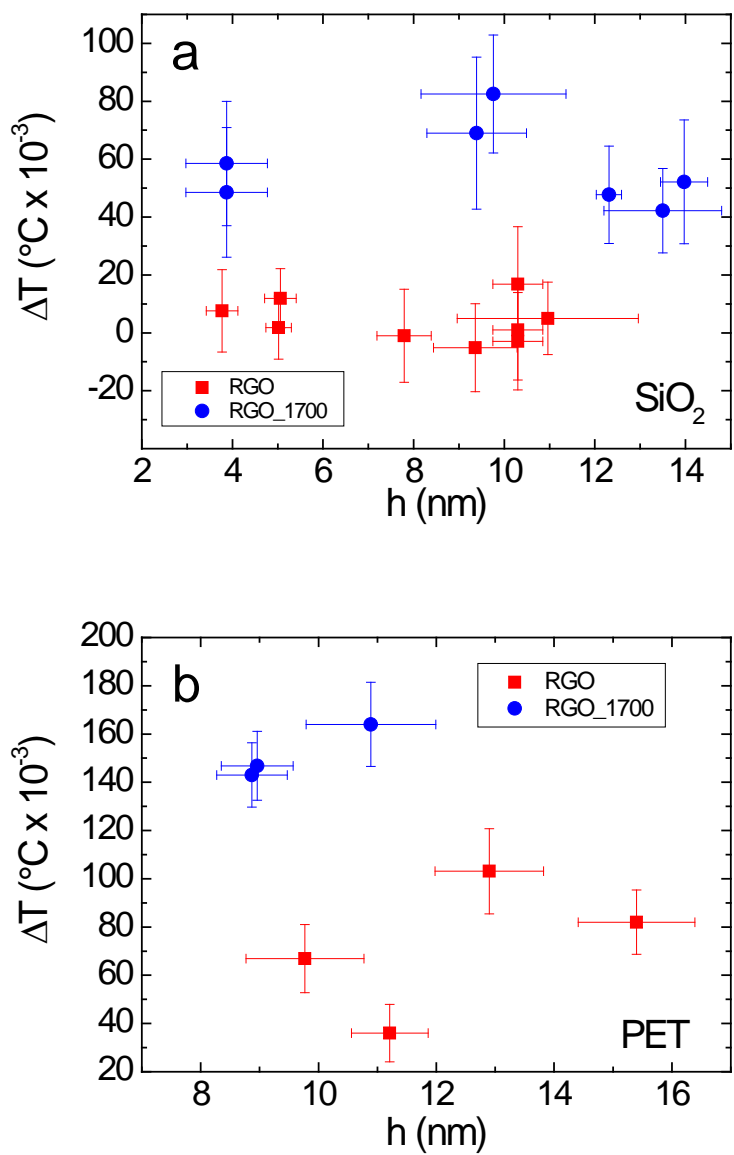

Figure 10: (a) Summary of the sensor temperature difference between the Si/SiO$_2$ substrate and the RGO (red) or RGO_1700 (blue) graphite nanoplates. (b) The same as in (a) but for the PET substrate.

A summary of the results obtained by using Si/SiO$_2$ as the substrate is reported in panel a of Figure 10. It can be seen that, although the temperature differences are small, the results are very well reproducible. Furthermore, it is worth noting that the temperature differences appear to be independent on the nanoplates thickness, in the thickness range (h ≈ 4-15 nm) explored in this work. This fact means that the variations in the explored thickness range do not produce an experimentally detectable change in the measured heater temperatures. This could be reasonable taking into account that the in-plane conductivity is expected to be substantially higher than the out of plane one and the heat flux could have a considerable component spread and carried away along the planes, thus limiting the effect of the variation of the flake thickness. The summary of the measurements carried out on flakes supported by PET is reported in Figure 10b. Based on the ΔT values obtained and their experimental deviations, also in this case, as observed for SiO$_2$/Si, no significant trend as a function of thickness can be claimed for their thermal response.

From the data reported in Figure 10, it is possible to calculate the average $\overline{\Delta T_{sub}} = \overline{T_{sub} - T_{flake}}$ values for the four cases we investigated (RGO and RGO_1700 on both substrates) and take the difference $\delta T_{sub} = \overline{\Delta T_{Sub}^{1700}} - \overline{\Delta T_{Sub}}$, where "sub" is either SiO$_2$ or PET. Thus, the quantity $\delta T_{sub}$ should give an idea, for each substrate, of the average temperature difference between the pristine and annealed RGO flakes, because $T_{sub}$ is constant for the same substrate material within the experimental uncertainty. We obtained $\delta T_{SiO_2} = (53.3 \pm 35.3) \times 10^{-3} K$ and $\delta T_{PET} = (79.3 \pm 29.3) \times 10^{-3} K$. The uncertainties for $\overline{\Delta T_{sub}}$ have been obtained by summing up the uncertainties of each term of the average and dividing by the number of terms, while for $\delta T_{sub}$ by summing up the uncertainty of each term of the subtraction. Although they are very large, the use of a low thermally conducting substrate like PET tends to enhance the experimental sensitivity for detecting the temperature difference between the two kinds of nanoplates, which is clearly related to their different thermal conductivity.

As it is possible to see from the results reported above, the temperature change that occurs when the tip passes from the substrate to the RGO or RGO_1700 samples is rather small (maximum ≈ 0.18 K). This means that the order of magnitude of the total "sample" resistances when the tip is on the RGO (i.e. the series of the nanoplate resistance with that of the substrate and of the interface) is approximately the same as of those reported above for the tip on the substrates.

As a further confirmation of this, theoretical calculations that take into account the anisotropic thermal conduction [46] of a nanoplate with dimensions similar to those measured here and values of the conductivities along the *ab* planes and the *c* axis within a realistic range, would give (only for the heat conduction along the *c* axis) a thermal resistance of the order of $10^5$ $K/W$, which is two orders of magnitude smaller than the total $R"$. It is therefore clear that, as observed, changes in the conductivity of the nanoplates produce small changes in the temperature of the sensor. Since besides their anisotropic conduction, the system is also considerably complicated by the small thickness of the nanoplates and the thermal interface with the substrate, as already remarked before, a detailed finite-element study of this problem should be performed in order to quantitatively determine the thermal conductivities of the samples.

## 4. Conclusions

Reduced graphite oxide nanoplates were studied in this paper as an interesting type of graphene-based material for the application in thermally conductive composites. In this field, the intrinsic thermal conductivity of individual RGO flakes is obviously crucial. Despite this, in composite literature such individual conductivity is typically not measured and sometimes assumed equal to values measured or calculated for graphene, neglecting the effects of flake thickness and the presence of defects. In this paper, a clear correlation between the defectiveness of RGO and its thermal properties was demonstrated by a thorough spectroscopic and thermal characterization of RGO both as obtained from thermal reduction of graphite oxide and after a high temperature annealing treatment. In particular, it was shown that a process of thermal annealing in vacuum for 1 h at 1700°C can considerably reduce the amount of defects in reduced graphite oxide nanoplates, as consistently proven by Raman measurements, X-ray photoelectron spectroscopy, X-ray diffraction and thermogravimetry. Thanks to this treatment, we were able to eliminate the majority of oxidized groups and to obtain a significant ordering of the structure both in plane and perpendicularly to the graphitic planes.

Scanning thermal microscopy images have been recorded on both pristine and annealed RGO flakes, simultaneously with the topography and lateral force maps thus combining in a single measurement properties analysis that cannot be observed at the same time with other techniques. In particular, the probes employed here allow obtaining topographic images and thermal properties with a spatial resolution of a few tens of nm.

The dramatic reduction of defects in annealed RGO comes along with a noticeable difference in temperature contrast between bare substrate and annealed RGO, compared to the case of pristine RGO. The increased temperature difference $T_{sub} - T_{flake}$ upon annealing evidences an increase in the thermal conductivity of the RGO, in qualitative agreement with the simple electric-analogue model with lumped parameters discussed in the paper. However, in order to address the extremely difficult task of thermal conductivity quantitative determination for the investigated samples, it would be necessary to properly model (e.g. by finite elements methods) the system. Such a complex model, which is beyond the scope of this paper, would require to take into account the SThM probe geometry and physical characteristics, the anisotropic thermal conductivity of the RGO flake along with the thermal properties of the supporting substrates and their interface interactions. Even without a comprehensive model, systems concerning with two different substrates hosting the RGO nanoplates clearly demonstrated that the temperature differences - and thus the sensitivity to the thermal conductance changes of the flakes - are increased when the nanoplates are supported by the less thermally conducting PET rather than $SiO_2$/Si.

The discussed results demonstrate that SThM is therefore a reliable tool to characterize in a rather simple and fast way the morphological and thermal properties of individual graphite nanoplates, multilayer graphene flakes or other 2D materials. This method was applied here to prove that the reduction of the defectiveness in graphene-based materials is indeed crucial to obtain the high thermal conductivity performance needed for the preparation of effective thermally conductive nanocomposites.

**Acknowledgements**

The research leading to these results has received funding from the European Union Seventh Framework Programme under grant agreement n°604391 Graphene Flagship. This work has received funding from the European Research Council (ERC) under the European Union's Horizon 2020 research and innovation programme  grant agreement 639495 — INTHERM — ERC-2014-STG. Funding from Graphene@PoliTo initiative of the Politecnico di Torino  is also acknowledged.

The authors gratefully acknowledge Dr. Mauro Raimondo and Salvatore Guastella at Politecnico di Torino-DISAT for FESEM and XPS analyses, respectively. E. Chiavazzo and A. Nastasi are gratefully acknowledged for their useful discussions.

**Authors contributions**

A. Fina conceived the experiments and coordinated the project, M. Tortello and R.S. Gonnelli designed and carried out SThM experiments and analyzed relative data. S. Colonna and M. Bernal analyzed characterization results. J. Gomez prepared pristine RGO and M. Pavese carried out thermal annealing of RGO. C. Novara and F. Giorgis carried out Raman measurements and their analyses. M. Maggio and G. Guerra carried out XRD measurements and their analyses. G. Saracco contributed to the discussion of the results. Manuscript was mainly written by M. Tortello and A. Fina.